\documentclass[conference]{IEEEtran}
\IEEEoverridecommandlockouts
\usepackage{cite}
\usepackage{amsmath,amssymb,amsfonts}
\usepackage{algorithmic}
\usepackage{graphicx}
\usepackage{textcomp}
\usepackage{xcolor}
\usepackage[most]{tcolorbox}
\usepackage{booktabs}
\usepackage{colortbl}
\usepackage{multirow}
\usepackage{url}
\usepackage{xurl}
\usepackage{bm}
\usepackage[justification=centering]{caption}
\usepackage{balance}
\usepackage{float}
\usepackage{comment}
\usepackage{hyperref}
\usepackage{fontawesome5}
\usepackage{balance}
\usepackage[normalem]{ulem}
\def\BibTeX{{\rm B\kern-.05em{\sc i\kern-.025em b}\kern-.08em
    T\kern-.1667em\lower.7ex\hbox{E}\kern-.125emX}}
\begin{document}

\title{How Do Developers Search for Architectural Information? An Industrial Survey}

\author{\IEEEauthorblockN{Musengamana Jean de Dieu and Peng Liang$^{*}$\thanks{\indent This work was funded by the National Key R\&D Program of China with Grant No. 2018YFB1402800 and the NSFC with Grant No. 62172311. The authors acknowledge the support from the China Scholarship Council.}}
\IEEEauthorblockA{School of Computer Science\\
Wuhan University, Wuhan, China\\
{\{jeanmusenga, liangp\}}@whu.edu.cn}
\and
\IEEEauthorblockN{Mojtaba Shahin}
\IEEEauthorblockA{Faculty of Information Technology\\
Monash University, Melbourne, Australia\\
mojtaba.shahin@monash.edu}
}

\maketitle

\begin{abstract}
Building software systems often requires knowledge and skills beyond what developers already possess. In such cases, developers have to leverage different sources of information to seek help. A growing number of researchers and practitioners have started investigating what programming-related information developers seek during software development. However, being a high level and a type of the most important development-related information, architectural information search activity is seldom explored. To fill this gap, we conducted an industrial survey completed by 103 participants to understand how developers search for architectural information to solve their architectural problems in development. Our main findings are: (1) searching for architectural information to learn about the pros and cons of certain architectural solutions (e.g., patterns, tactics) and to make an architecture decision among multiple choices are the most frequent purposes or tasks; (2) developers find difficulties mostly in getting relevant architectural information for addressing quality concerns and making design decisions among multiple choices when seeking architectural information; (3) taking too much time to go through architectural information retrieved from various sources and feeling overwhelmed due to the dispersion and abundance of architectural information in various sources are the top two major challenges developers face when searching for architectural information. Our findings (1) provide researchers with future directions, such as the design and development of approaches and tools for searching architectural information from multiple sources, and (2) can be used to provide guidelines for practitioners to refer to when seeking architectural information and providing architectural information that could be considered useful.
\end{abstract}


\section{Introduction} \label{Introduction}
Software developers often leverage various sources (online and offline) of software development knowledge to seek information in order to support development activities \cite{rahman2018evaluating}, \cite{gu2016deep}, \cite{stolee2014solving}, \cite{sim2013controlled}, \cite{sim2011well}. Developers may search for development information for different purposes or tasks during various phases of software development. For instance, during the implementation phase, developers may look for reusable code snippets, suitable third-party libraries, or the usage of particular APIs \cite{xia2017developers}. During the maintenance phase, developers may seek solutions to address bugs \cite{ko2006exploratory}. A number of empirical studies have extensively investigated how developers perform code search to support implementation phase tasks (e.g., \cite{sim1998archetypal}, \cite{sadowski2015developers}, \cite{stolee2014solving}, \cite{brandt2009two}). Sim \textit{et al}. \cite{sim1998archetypal} found that the most common reasons for code search include defect repair, code reuse, program understanding, impact analysis, and feature addition. Sadowski \textit{et al}. \cite{sadowski2015developers} conducted a study at Google to investigate how developers search for code, in what contexts are code search tools used, and what are the properties of search queries. In a survey, Stolee \textit{et al}. \cite{stolee2014solving} found that 85\% of developers search the web for source code at least weekly.

\begin{figure}[h]
  \includegraphics [scale=0.38] {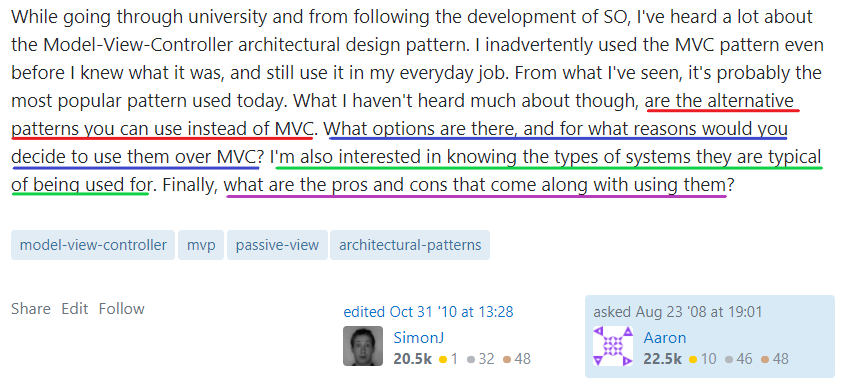}
 \caption{A Stack Overflow post in which a user was seeking architectural information about the MVC pattern}
 \label{searchExample}
\end{figure}

Prior work has mainly focused on how developers perform code search to support development tasks. Yet, code search is only one of the search tasks that developers perform in development. Developers search for many other tasks or purposes, such as searching for the use of architectural patterns, including their benefits and drawbacks, to choose a solution for an architectural problem. Figure \ref{searchExample} shows a Stack Overflow question\footnote{\url{https://tinyurl.com/2d8r6w8m}}, in which a user was searching for architectural information about \textit{alternative architectural patterns for Model View Controller (MVC) pattern} (i.e., alternative architecture solutions \cite{zimmermann2009managing}). In the question body, the user sought \textit{the reasons that could drive someone to decide to use them over MVC} (i.e., architecture decisions and their rationale \cite{jansen2005software}), \textit{the types of systems they are typical of being used for}, and \textit{the pros and cons that come along with using them} (i.e., benefits and drawbacks of architecture solutions \cite{soliman2015enriching}).

A comprehensive understanding of how developers search for architectural information can help practitioners better understand the frequent purposes and tasks of searching architectural information. It can also help researchers be aware of difficulties and challenges developers are facing when seeking architectural information. Thus, architecture communities could design dedicated approaches and tools to assist developers to effectively and easily search and find relevant architectural information according to their development tasks or purposes. Hence, in this study, we conducted an industrial survey \cite{personal2005} to investigate the architectural information search activities of developers in their daily work. The survey includes both closed-ended and open-ended questions and was completed by 103 participants across 6 continents (i.e., 40 countries).

The main findings of this survey are: (1) search engines (e.g., Google), online Q\&A sites (e.g., Stack Overflow), and online repositories (e.g., GitHub) are the top three most frequently leveraged sources to seek architectural information, (2) seeking architectural information to learn about the pros and cons of certain architectural solutions (e.g., patterns, tactics) and to make an architecture decision among multiple choices are the most frequent purposes or tasks, (3) it is most difficult for developers to seek and find relevant architectural information to address quality concerns and make design decisions among multiple choices when searching for architectural information, (4) dedicated approaches and tools for searching architectural information (proposed by researchers in the literature) are not adopted in practices, and (5) spending too much time to review architectural information retrieved from various sources and feeling overwhelmed due to the dispersion and abundance of architectural information in various sources were reported as the top two major challenges that the participants faced when seeking architectural information.

The remainder of this paper is structured as follows. Section \ref{relatedwork} summarizes related work. Section \ref{StudyDesign} describes the survey design. Section \ref{surveyResults} presents the survey results. Section \ref{Discussion} presents the key findings with implications of each RQ. Section \ref{Threats to Validity} highlights the threats to validity. Section \ref{Conclusions and Future Work} concludes this survey with potential areas of future research.

\section{Related Work} \label{relatedwork}

\subsection{Code Search}
Uddin \textit{et al}. \cite{uddin2019understanding} conducted two surveys with 178 developers to understand how developers seek and evaluate API opinions. They observed that developers seek API reviews and often have to summarize those APIs reviews for diverse development needs (e.g., API suitability). In a survey with 69 developers, Sim \textit{et al}. \cite{sim1998archetypal} found that the most common motivators behind code search are code reuse, defect repair, program understanding, feature addition, and impact analysis. Stolee \textit{et al}. \cite{stolee2014solving} found that 59\% of the surveyed developers perform code search daily. Robillard and DeLine conducted a survey and a series of qualitative interviews of developers at Microsoft to understand how developers learn APIs. They found that the most severe obstacles developers faced while learning new APIs were related to the official documentation of the APIs (e.g., the lack of code examples and the absence of task-oriented description of the API usage) \cite{robillard2011field}. Umarji \textit{et al}. \cite{umarji2008archetypal} investigated developers’ code searching practices using a questionnaire, and found that the factors affecting the final selection of a candidate piece of code include peer recommendations, availability of help from other programmers, easy of installation, and availability of documentation.

Search logs are also a good dataset to study the search behavior of developers. By analyzing search logs, Brandt \textit{et al}. \cite{brandt2009two} found that 48\% of the queries contain just code, 38\% contain just natural language, and 14\% contain a mix of both. Bajracharya and Lopes conducted an extensive analysis of a log file of an Internet-scale source code search engine, Koders, to categorize the topics that provide insights on the different ways code search engine users express their queries \cite{bajracharya2012analyzing}. Masudur \textit{et al}. \cite{rahman2018evaluating} collected search logs from 310 developers that contain nearly 150,000 search queries from Google to explore whether a general-purpose search engine like Google is an optimal choice for code-related searches. They observed that code-related search often requires more effort (e.g., time and query modifications) than general non-code search.

These studies mainly focus on code search and reuse practice to solve implementation tasks or problems in development. To the best of our knowledge, no studies investigate how developers seek architectural information to solve architecture-related tasks or problems. Therefore, our study differs from the existing studies in that it offers an investigation of architectural information search activities, including sources used to look for architectural information, tasks or purposes of the search, methods and tools utilized, assessment of the usefulness of retrieved architectural information, and the challenges and issues faced when searching for architectural information.

\subsection{Approaches and Tools for Architectural Information Search} \label{MiningArchInfo}
A few approaches and tools have been reported to support searching architectural information. Bhat \textit{et al}. \cite{bhat2017automatic} proposed an approach to automatically identify and classify design decisions in issue management systems. van der Ven\textit{et al}. \cite{van2013making} proposed an approach to automatically analyze and capture information about design decisions from large open-source software repositories. A domain-specific automatic search approach has been proposed in \cite{soliman2018improving} to improve the search of architectural information in Stack Overflow. Bi \textit{et al}. \cite{bi2021mining} developed a semi-automatic approach to mine posts from Stack Overflow and structured the design relationships between architectural tactics and quality attributes used in practice to help architects better make design decisions. Shahbazian \textit{et al}. \cite{shahbazian2018recovering} developed a technique called RecovAr to automatically recover design decisions from the readily available history artifacts of projects (e.g., an issue tracker and version control repository). In their follow-up work, they developed a tool named ADeX for the automatic curation of design decisions from Jira, GitHub, MS project, and Enterprise Architect, to enrich and generate specific views on architectural knowledge, which consequently supports developers and architects’ decisions making process \cite{bhat2019adex}. 

Although the abovementioned approaches and tools have been proposed to search architectural information, none has been studied from the practitioners’ perspectives. Thus, one of the motivators of our industrial survey presented in this paper is to examine whether or not practitioners have adopted the above discussed dedicated approaches and tools in practice.

\section{Study Design} \label{StudyDesign}

\subsection{Goal and Research Questions} \label{GoalRearchQuestions}
The main \textbf{goal} of this study defined using the Goal-Question-Metric (GQM) approach [11] is: \textit{to \textbf{analyze} the current situation on seeking architectural information \textbf{for the purpose} of understanding \textbf{with respect to} the available sources of architectural information, purposes and tasks searched, methods (manual, semi-automatic, automatic) and tools employed, usefulness of retrieved information, and issues \textbf{from the point of view of} developers (practitioners) \textbf{in the context of} software development}. Following the goal of this study, we formulated five Research Questions (RQs).

\textbf{RQ1. What are the sources that developers use to search for architectural information?}

\textbf{Rationale}: Architectural information is scattered in various sources, such as Q\&A sites \cite{soliman2016architectural}, issues tracking systems \cite{bhat2017automatic}, and technical blogs and tutorials \cite{soliman2021explorin}. Each source contains different types of architectural information (e.g., benefits and drawbacks of architectural solutions) \cite{soliman2021explorin}. Thus, there is no single source that contains all required architectural information. Depending on the type of architectural information, one may need to look into multiple source of architectural information. The answer to this RQ provides insight about specific sources developers (practitioners) leverage when searching for architectural information. Such insight can help (i) researchers know if a certain source is more used than other sources so that the analysis of the source can be given more priority over the others, and (ii) developers to determine the scope of searching (i.e., to search in specific sites or search in the whole web).

\textbf{RQ2. What are the purposes or tasks that require developers to search for architectural information?}

\textit{\textbf{Rationale}}: Developers may seek architectural information for various purposes or tasks, such as making an architectural decision among multiple choices. The answer to this RQ can provide researchers with an overview of the purposes or tasks that require developers to search for architectural information, as well as the tasks or purposes that developers find difficult to get relevant architectural information. This can further motivate researchers to address the difficulties in searching relevant architectural information.

\textbf{RQ3. What methods (manual, semi-automatic, automatic) and tools do developers use to search for architectural information?}

\textbf{Rationale}: Various methods and tools have been used by developers to search for architectural information according to their needs and preferences. The answer to this RQ can not only let practitioners be aware of these approaches and tools, but also help researchers identify the gap between academia and industry in the adoption of the approaches and tools.

\textbf{RQ4. How do developers assess the usefulness of retrieved architectural information?}

\textbf{Rationale}: The usefulness of retrieved architectural information is vital for the purposes or tasks that require architectural information. The answer to this RQ aims to provide information about what factors and measures are employed by developers when assessing the usefulness of retrieved architectural information, which can further help practitioners gain insights into the usefulness measures or factors to rely on when providing architectural information that can be considered useful, and motivate researchers to investigate new factors and measures or approaches and tools for assessing the usefulness of architectural information.

\textbf{RQ5. What issues do developers face when searching for architectural information?}

\textbf{Rationale}: There might be issues and challenges that developers are facing when looking for architectural information. The answer to this RQ provides information about what significant issues developers face when looking for architectural information. By learning about these issues, researchers can gain insights into the complexity of the problems that need to be addressed and assist developers when searching for architectural information.

\subsection{Survey Design}\label{SurveyDesign}
We conducted a descriptive survey \cite{personal2005} by following the guidelines proposed by Kitchenham and Pfleeger \cite{personal2005}. The survey questionnaire has been made available online\footnote{\url{https://tinyurl.com/yckunkn6}}.

\subsubsection{The structure of the survey questionnaire}\label{Creating the questionnaire}
Before creating our survey questionnaire, a thorough review of the related work to our study was performed. When creating our survey questionnaire, we ensured that the Survey Questions (SQs) were linked to our main RQs. Briefly, the survey starts with an Introduction page that explains the goal of the survey along with some preliminary definitions and the time frame (i.e., 10 to 15 minutes) that a participant could take to finish the survey. The first part of the survey includes demographic questions (i.e., SQ1-SQ5). For example, we asked the role of the participants (e.g., developer, architect, tester) and their years of experience in software development. Such data can help us to assess the participants’ qualification for the survey based on the inclusion and exclusion criteria defined in Table \ref{InclusionAndExcuOfvalidResponses}. To be more specific, for the inclusion criterion, we considered the participants that have more than two years of experience in development because we did not want to report biased results due to, for example, a lack of experience and expertise.

\begin{table} [h] 
\small
	\caption{Inclusion and exclusion criteria for filtering valid responses}
	\label{InclusionAndExcuOfvalidResponses}
	\begin{tabular}{p{8.2cm}}
		\toprule
	\textbf{Inclusion criterion}
		\\ \midrule
	\textbf{I1:} The participant has more than two years of experience in software development.
	\\ \midrule
	\textbf {Exclusion criterion}
	\\ \midrule
    \textbf{E1:} The response is nonsense or randomly filled.
	\\	\bottomrule 
	\end{tabular}
\end{table}

The remaining parts of the survey are (1) questions on how and where developers seek architectural information (i.e., SQ7-SQ10), (2) questions on the purposes or tasks that motivate developers to look for architectural information (i.e., SQ11 and SQ12), (3) questions on the methods or tools developers use (i.e., SQ13(a)-SQ13(d)), (4) a question on how the usefulness of retrieved architectural information is assessed (i.e., SQ14), and (5) a question on the issues/challenges faced when seeking architectural information (i.e., SQ15). 
In total, our survey is composed of 16 questions, which are all mandatory except for two questions (i.e., SQ6 and SQ16). SQ6 asks the participant's email address for sharing the survey results in case s/he wants them, and SQ16 asks for comments or remarks about relevant aspects (if any) on searching architectural information that is possibly left uncovered. 

Our survey is mixed with close-ended questions (e.g., SQ3, SQ17, SQ8, SQ9(a), SQ11) and open-ended questions (i.e., SQ9(b), SQ13(a), SQ13(b), SQ13(c), SQ13(d)). The close-ended questions are composed of multiple-choice, and they include numerical values, Yes/No answers, and Likert scale questions \cite{likert1932technique}. In closed-ended questions, we provided several answer options for the participants to choose from, but we also provided an open textual field, i.e., “Other", allowing participants to add their own responses freely. We adopted this approach of providing an open textual field to closed-ended questions as we did not want to impose restrictions and avoid bias on the participants' responses.

\subsubsection{Target population recruitment and sampling}\label{Targe population}
We used several approaches to reach the target population. Firstly, similar to the previous studies \cite{waseem2021design}, \cite{tian2022relationships}, and \cite{uddin2019understanding}, we used GitHub developers by collecting their email addresses provided in users’ profiles. The reason we selected GitHub for our survey is that it is a popular Open-Source Software (OSS) repository. In addition, previous research shows that GitHub developers use various sources (e.g., online Q\&A sites) to look for development-related information (e.g., seeking opinions about APIs \cite{uddin2019understanding}) during their OSS development. Secondly, we posted our survey in the groups of software engineers on social media sites (e.g., LinkedIn\footnote{\url{https://www.linkedin.com/groups/6608681/}} and Facebook\footnote{\url{https://www.facebook.com/groups/696598770719379/}}) where practitioners from different countries share their issues, experiences, and knowledge in software development. Finally, to increase the response rate of our survey, we sent reminder emails and utilized snowballing sampling \cite{kitchenham2002principles} to request our potential participants to invite their colleagues who might want to take part in our industrial survey. 

 \subsubsection{Validation and evaluation of the survey design}\label{Evaluating and validating the questionnaire} 
Before hosting the survey online for data collection, all authors thoroughly reviewed the survey design internally by cross-checking several parts in the survey protocol, such as survey objectives, survey questions, and RQs. Moreover, we also performed a pilot survey by sending invitations to five participants through their email addresses that we gathered from GitHub. Among the five participants, three are team leaders, and two are professional developers. All five participants are actively involved in software development with 3 to 10 years of experience. During the pilot survey study, we examined various factors that could influence the outcomes of the survey, such as the comprehensibility of the survey questions and the time frame needed to answer the questions. Based on the responses from the pilot survey study, we refined one survey question (i.e., SQ13) to make it more clear and understandable.
 
 \subsubsection{Filtering valid responses from the formal survey}
After sending the invitations through the gathered emails and groups of software engineers on social media sites, we obtained 117 responses. Among those, we excluded 14 responses based on the inclusion and exclusion criteria defined in Table \ref{InclusionAndExcuOfvalidResponses}. We finally got 103 valid responses.

\subsection{Survey Data Analysis}\label{SurveyDataAnalysis}

We analysed the survey responses using two approaches. More specifically, the descriptive statistics \cite{wohlin2003empirical} approach was used for better understanding the occurrences of each given response to each closed-ended question (see Table \ref{dataAnaysisMethodandRQ}). Open coding and constant comparison techniques from Grounded Theory (GT) \cite{stol2016grounded} were used to analyze the responses to open-ended questions (see Table \ref{dataAnaysisMethodandRQ}). We utilized a qualitative data analysis tool (MAXQDA\footnote{https://www.maxqda.com/}) to support the analysis process. Specifically, the first author (i.e., a PhD student) summarized the main idea stated in each response of each open-ended question. He went on to label the summarized idea (from each response) to generate codes. For example, a participant reported why (reason) s/he seeks architectural information in online repositories, such as GitHub (i.e., Q9(b)) by saying: “\textit{I am a developer, and I want to see how discussed design information was implemented in the code. So, GitHub holds my expectations}". In this case, we coded that response as “architectural information implemented in code".  

Moreover, in the open-ended questions, we encoded a response as “\textit{Not sure}" when the participant in his or her response mentioned that s/he was not sure of the specific answer. For example, the following responses were all coded as “\textit{Not sure}": (1) “\textit{I don’t know}”, (2) “\textit{I would not think anything}", and (3) “\textit{I am not sure}”. Afterwards, the first author applied constant comparison to compare the code identified in one summarized idea with the codes that emerged from other summarized ideas to see the codes which have similar semantic meanings. He proceeded to group similar codes into high-level categories. To mitigate the personal bias during data analysis, the other authors of this study participated in the validation of the generated codes. In the case where there was any disagreement, all authors discussed and resolved it until a consensus was reached. The survey results are reported in Section \ref{surveyResults}. We also provided the valid survey responses in MS Excel and the encoded data in MAXQDA online~\cite{dataset}.

\begin{table} [h!]
\small
\caption{Data analysis approaches used to analyze survey question data and their corresponding RQs} \label{dataAnaysisMethodandRQ}
	\begin{tabular}{p{2.5 cm}p{2.7cm}p{2cm}}
		\toprule
	    Survey question & Data analysis approach & RQs\\
		\midrule
	   SQ1-SQ6 & Descriptive Statistics & Demographics\\
		
	   SQ7, SQ8, SQ9(a), SQ10 & Descriptive Statistics & RQ1\\
		
	   SQ11, SQ12 & Descriptive Statistics & RQ2\\
		
	   SQ9(b), SQ13 & Open coding \& constant comparison & RQ1 and RQ3 \\
	   SQ14-SQ15 & Descriptive Statistics & RQ4 and RQ5\\
		\bottomrule 
	\end{tabular}
\end{table}

\section{Survey Results}\label{surveyResults}

\subsection{Demographics}\label{Demographics}
\textbf{Country}: The survey participants come from 6 continents (40 countries). Most participants come from the US (17.4\%, 18 out of 103), followed by India (15.5\%, 16 out of 103). 

\textbf{Role}: Our participants cover different roles, such as project manager, team leader, architect, developer, and requirements engineer. A large number of participants worked as software developer (42.1\%, 48 out of 103), followed by architect (18.5\%, 21 out 103) and team reader (16.7\%, 19 out of 103). 

\textbf{Experience}: The distribution of experience of the participants is: 20+ years (35 out of 103, 33.9\%), 10 to 20 years (31 out of 103, 30.1\%), 4 to 10 years (25 out of 103, 24.2\%), and 2 to 4 years (12 out of 103, 11.6\%). 

\textbf{Project size}: The participants reported different sizes of the largest projects they had worked on (number of people involved). A majority of them, i.e., 20.3\% (21 out 103), worked in the projects where people ranged between 50-100. 

\textbf{Organization domain}: The domains of the organizations where the participants worked in vary significantly, including consulting and IT services, embedded systems, E-commerce, financial, healthcare, telecommunication, retail, and insurance. Most of them worked in consulting and IT service domain (59.2\%, 61 out of 103).

A diversified population of participants gives us confidence that our recruited participants are representative of the survey target population.


\subsection{RQ1: How and where developers search for architectural information}\label{RS1_Results}
We explored this RQ by asking participants five survey questions (SQ7-SQ10). Firstly, we asked the participants if they search for architectural information in software development (i.e., SQ7). All the participants responded that they do search for architectural information in development. The results of the remaining four survey questions (SQ8, SQ9(a), SQ9(b), and SQ10) are described bellow.  

\subsubsection{Sources of architectural information} 
\begin{figure}[h]
  \includegraphics [scale=0.38] {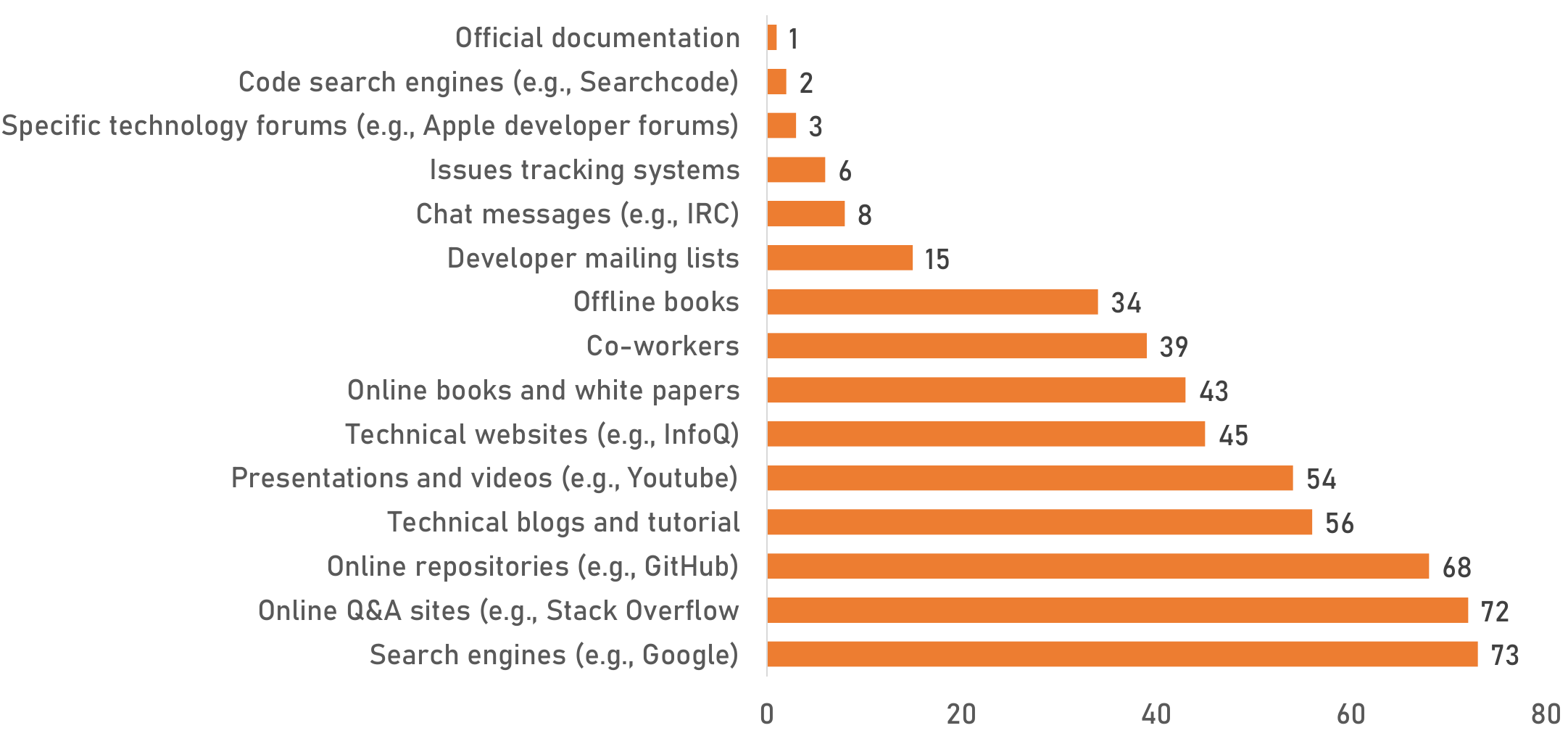}
 \caption{Fifteen types of sources that are visited to seek architectural information}
 \label{SourcesAndParticipants}
\end{figure}

We asked the participants where (sources) they seek architectural information in development (i.e., SQ8). The participants were given thirteen options (i.e., categories of the sources) to choose from. However, as mentioned in the above sections, we provided an open textual field, i.e., “Other", to let participants provide their own relevant answers (sources). In the open textual field, we collected the sources, such as SourceForge (reported 2 times), Searchcode (reported 1 time), Apple developer forums (reported 1 time), official documentations (reported 1 time). We observed that most of the mentioned sources as “Other” fall into our provided categories of the sources. For example, SourceForge belongs to “online repositories" category. We added categories of sources that do not fall into any of our provided source categories. For example, “code search engines (e.g., Searchcode)" and “specific technology forums (e.g., Apple developer forums)" were added as new categories of sources to our provided list. As shown in Figure \ref{SourcesAndParticipants}, the most frequently visited source when searching for architectural information is search engines (e.g., Google) (reported 73 times), followed closely by online Q\&A sites (e.g., Stack Overflow) (reported 72 times), and online repositories (e.g., GitHub) (reported 68 times). Note that, since a participant was able to provide more than one response (source), the sum of the sources reported in Figure \ref{SourcesAndParticipants} exceeds the total number of the survey participants. 

\subsubsection{The most useful sources of architectural information} 
\begin{figure}[h]
  \includegraphics [scale=0.40] {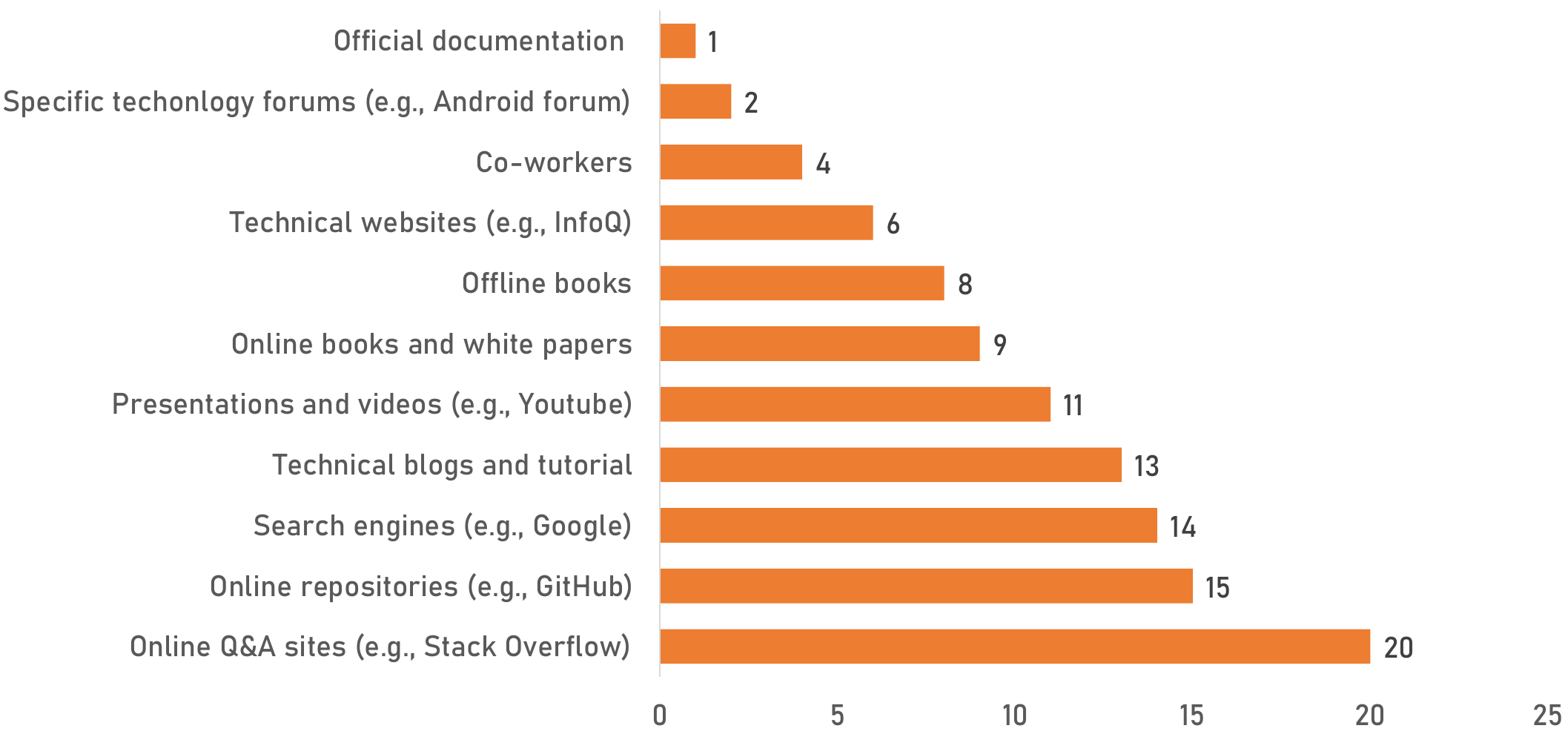}
 \caption{Eleven types of sources that are considered the most useful when seeking architectural information}
 \label{MostUsefulSources}
\end{figure}

Once the participants provided the source(s) they visit to look for architectural information to answer SQ8, we further asked them which source is the most useful when searching for architectural information (i.e., SQ9(a)). From the responses to SQ9(a), the participants reported eleven types of sources they consider to be the most useful when searching for architectural information (see Figure \ref{MostUsefulSources}). As shown in Figure \ref{MostUsefulSources}, online Q\&A sites (e.g., Stack Overflow) (reported 20 times), online repositories (e.g., GitHub) (reported 15 times), search engines (e.g., Google) (reported 14 times), and technical blogs \& tutorials (reported 13 times) are the top four sources of architectural information that are considered the most useful when seeking architectural information. 

\subsubsection{The reasons for a source being useful} Through an open-ended question, we asked the participants why the source that they mentioned in SQ9(a) is the most useful when searching for architectural information (i.e., SQ9(b)). We have collected 33 reasons that explain why such a given source is the most useful over others from the participants' perspective. Due to the space limit, we only present the reasons reported on the top three sources (see Figure \ref{MostUsefulSources}), and the complete list of sources and reported reasons on each source can be found in our dataset \cite{dataset}.

\textbf{Online Q\&A sites}: 19.4\% (20 out of 103) participants (see Figure \ref{MostUsefulSources}) reported to find online Q\&A sites (e.g., Stack Overflow) the most useful when seeking architectural information with the following reasons:

\dashuline{(i) Source of solutions to problems faced in real world}: 14.5\% (15 out of 103) participants mentioned online Q\&A sites to be the most useful when searching for architecture information because they can get solutions to architectural problems faced in real-world scenarios. For example, one participant stated: “\textit{That's where I get most of the answers to architecture problems faced by my fellow developers in real-world scenarios}" \textbf{P39}.

\dashuline{(ii) Different opinions for design decisions}: 4.8\%, (5 out of 103) participants thought that the quality of architectural information can vary considerably. Moreover, not all architectural information may suit the needs at hand. Thus, they anticipated seeing the benefits and drawbacks of certain architectural solutions (e.g., frameworks or architectural patterns) in online Q\&A sites to make a design decision. One participant pointed out: “\textit{it's just that I find various discussions, such as advantages and disadvantages of certain frameworks, to help me decide on which solution is the best according to my design problem. So that information is, in general, of reasonably good quality}" \textbf{P87}.

\textbf{Online repositories}: 14.5\% (15 out of 103) participants (see Figure \ref{MostUsefulSources}) cited the following reasons that motivate them to seek architectural information in online repositories (e.g., GitHub):

\dashuline{(i) Architectural information implemented in code}: 7.7\% (8 out of 103) participants said that they directly visit online repositories (e.g., GitHub) when looking for architectural information because they can get detailed information in both source code files \cite{gopalakrishnan2017can} and texts (e.g., through architecture design discussions in pull requests \cite{viviani2018structure}, commits and issues \cite{brunet2014developers}). For example, “\textit{I am a developer, and I want to see how discussed design information was implemented in the code. So, GitHub holds my expectations}" \textbf{P26}.

\dashuline{(ii) Understanding the architecture through the code}: 6.7\% (7 out of 103) participants chose to visit online repositories over other sources since online repositories help them to perceive the architectures/structures of the projects through the implemented code. For example, “\textit{...General, I can understand the structure of the project being implemented (in code) although I have not participated before}" \textbf{P82}.

\textbf{Search engines}: 13.5\% (14 out 103) participants mentioned search engines as the most useful sources when seeking architectural information for the following reasons:

\dashuline{(i) Source of diverse architectural information in various formats}: Some participants (5.8\%, 6 out of 103) indicated that search engines guide them to retrieve different kinds of architectural information (e.g., architectural tactics, etc.) in different forms (e.g., textual, diagrams, videos, etc.). For example, “\textit{To me, Google plays as a source of different architectural contents in various formats}" \textbf{P22}.

\dashuline{(ii) Serve as gateways to sites with architectural content}: 4.8\%, 5 out of 103 participants mentioned that search engines act as gateways to other sites discussing software architecture. For instance, “\textit{It (Google) is a search engine which serves as a gateway or middleman between me and my answers on the Internet, it leads me to other sites with contents related to software design}" \textbf{P99}.

\subsubsection{The frequency of searching for architectural information} We also asked the participants how often they seek architectural information in software development (SQ10). There were four options to choose from. 37.8\% (39 out of 103) participants mentioned that “depend on the tasks in hand", 33.9\% (35 out 103) participants search for architectural information “several times in a week", and 15.5 \% (16 out 103) participants look for architectural information “several times in a month". Only 12.6\% (13 out of 103) participants mentioned that they search for architectural information “every day".



\noindent\resizebox{\columnwidth}{!}{\fbox{%
	\parbox{\columnwidth}{%
		\textbf{Key Findings of RQ1}
		\\
\textbf{Finding 1}: Search engines (e.g., Google), online Q\&A sites (e.g., Stack Overflow), and online repositories (e.g., GitHub) are respectively the top three most frequently visited sources to seek architectural information. 

\textbf{Finding 2}: Online Q\&A sites (e.g., Stack Overflow) and online repositories (e.g., GitHub) are the top two sources that are considered the most useful when searching for architectural information. 
	}
}}


\subsection{RQ2: Purposes or tasks for searching for architectural information}\label{RQ2_Results}
To investigate this RQ, we first asked the participants for what purposes or tasks do they search for architectural information (SQ11), and the results of this survey question are presented in Table \ref{PurposesOfsearchingArchInfo}. Second, we asked them to cite the tasks or purposes they find difficult to get relevant architectural information (SQ12), and the results of this survey question are listed in Table \ref{TasksDifficTofind}. Table \ref{PurposesOfsearchingArchInfo} shows that participants seek architectural information to support diverse architectural design tasks or purposes in software development, such as learning about the impact of certain technologies on quality attributes, making an architecture decision among multiple choices, etc. On the other hand, Table \ref{TasksDifficTofind} shows that participants find difficulties mostly in getting relevant architectural information for addressing quality concerns and making design decisions among multiple choices when seeking architectural information.
 
\begin{table}[h]
\small
\caption{Purposes or tasks that motivate developers to search for architectural information} \label{PurposesOfsearchingArchInfo}
	\begin{tabular}{p{7.3cm}p{0.6cm}}
		\toprule
	    Purposes or tasks & Count\\
		\midrule
	  Learn about the pros and cons of certain architectural solutions (e.g., patterns, tactics) & 70\\
      Make an architecture decision among multiple choices & 68\\
	  Provide architectural information for addressing functional requirements &	64\\
      Learn about the implementation of certain architectural solutions (e.g., patterns, tactics) &54\\
      Provide architectural information that contains architectural decisions in a similar context&	49\\
      Provide architecture information for addressing quality concerns&	48\\
      Learn about architectural concepts or tools	& 46\\
      Learn about the existing architectures of systems &	46\\
      Learn about the impact of certain technologies on quality attributes&	42\\
      Improve architecture documentation & 27\\
		\bottomrule 
	\end{tabular}
\end{table}

\begin{table}[h]
\small
\caption{Purposes or tasks that are difficult to get relevant architectural information} \label{TasksDifficTofind}
	\begin{tabular}{p{7.3cm}p{0.6cm}}
		\toprule
	    Purposes or tasks & Count\\
		\midrule
     Provide architecture information for addressing quality concerns & 46\\
     Make an architecture decision among multiple choices &	44\\
     Provide architectural information for addressing functional requirements & 40\\
     Learn about the impact of certain technologies on quality attributes & 35\\
     Provide architectural information that contains architectural decisions in a similar context & 35\\
     Learn about the pros and cons of certain architectural solutions (e.g., patterns, tactics) & 33\\
     Learn about the implementation of certain architectural solutions (e.g., patterns, tactics) & 25\\
     Improve architecture documentation &	17\\
     Learn about the existing architectures of systems &	17\\
     Learn about architectural concepts or tools	& 14\\
     None                                           &2\\
	\bottomrule 
	\end{tabular}
\end{table}

\noindent\resizebox{\columnwidth}{!}{\fbox{%
	\parbox{\columnwidth}{%
		\textbf{Key Findings of RQ2}
		\\
\textbf{Finding 3}: Seeking architectural information to learn about the pros and cons of certain architectural solutions (e.g., patterns, tactics) and to make an architecture decision among multiple choices are the top two most frequently searched tasks or purposes.

\textbf{Finding 4}: Most participants find difficulties in getting relevant architectural information for addressing quality concerns, followed by identifying relevant architectural information for making design decisions among multiple choices.
	}
}}

\subsection{RQ3: Methods and tools used to search for architectural information}\label{RQ3_Results}
The participants were asked how they search for architectural information (SQ13(a)) and what dedicated method(s) (manual, semi-automatic, automatic) (SQ13(b)) and tool(s) (SQ13(c)) they use to seek architectural information.  

Most of the participants, i.e., 79.6\% (82 out of 103), reported that they use the automatic searching approaches provided in the existing tools, such as general-purpose search engines (e.g., Google) and Q\&A sites (e.g., Stack Overflow). They leveraged these tools by using various searching mechanisms, such as “\textit{keyword search with search engines}", “\textit{keyword search with search engines and the limit to specific sites}". For example, when searching with Google, one participant stated that: “\textit{I use phrases such as `comparison' or `evaluation' as well as constraining to a particular site with `site:domain.org'}" \textbf{P1}. Also, several participants mentioned that they use “\textit{keyword search in specific sites}" instead of starting their searches from general-purpose search engines.
 
Other (i.e., 12.6\%, 13 out of 103) participants stated that they use manual methods (e.g., offline book reading). 7.7\% (8 out of 103) participants mentioned that they utilize semi-automatic methods (e.g., using Web crawlers to mine architectural information from the web and manually check the relevant information). None of the participants reported using any dedicated tool(s) (SQ13(c)) or approach(es) (SQ13(b)) (proposed in the literature as described in Section \ref{MiningArchInfo}) to seek architectural information.

Moreover, we asked the participants if they believe that there is a need for such tool(s) that can help searching architectural information (SQ13(d)). If the answer is “Yes", we further asked the participants to provide the top three things they need/would like to see in such tool(s) (SQ13(d)). 72.8\% (75 out of 103) participants replied “Yes", 15.5\% (16 out of 103) participants said “No", and 11.6\% (12 out of 103) participants responded “Not sure". The participants who replied “Yes" provided their opinions about the top three requirements they would like to see in such tool(s). A significant example from the participants who replied "Yes" is “\textit{a dedicated search engine capable of indexing architectural information as its main (or only) goal and adding the kind of features that are supported by other searching engines (say, Manticore/Sphinx, or ElasticSearch...) would be a huge step ahead in terms of finding relevant information quickly}". We qualitatively analysed the provided responses and categorized them into 17 categories of the future tool support requirements. Due to the space limit, we only report the top five most cited tool support requirements in the following, and the complete list of tool support requirements can be found in our dataset \cite{dataset}. 

\dashuline{(i) Concise, relevant, and readable information}: 20.0\% of the participants who replied “Yes" (i.e., 15 out 75) suggested this opinion for future tool designs. Such tool(s) should provide brief but comprehensive, relevant, and readable architectural information of the searched task(s). One typical example stated by a developer is “\textit{the tool should focus on the relevancy between the query and the results. It should briefly summarize the contents and bring readable design information}" \textbf{P83}.

\dashuline{(ii) Architecture information with real world examples} was also recognized as an important requirement for future tool design by 16.0\% (12 out 75) participants. For example, one participant mentioned “\textit{I would like to be provided the following: A solid technical example of the implementation of the architecture in the relevant programming language}" \textbf{P27}.

\dashuline{(iii) Comprehensive pros and cons of architectural information}: 13.3\% (10 out 75) participants want a tool that can automatically highlight the benefits and drawbacks of retrieved architectural information (e.g., information about the pros and cons of certain technologies in a particular application domain), which can help to carefully make a sensible design decision. For example, one participant expected that “\textit{I would say comprehensive comparisons between pros \& cons of using such technologies in the architecture (...)}" \textbf{P66}.

\dashuline{(iv) {Capable of indexing relevant architectural information}} is also perceived as a significant requirement for future tool(s) by 10.6\% (8 out 75) participants.

\dashuline{(v) {Performance in retrieving architectural information}}: 9.3\% (7 out 75) participants highlighted the importance of considering the speed or performance in retrieving architectural information for future tool requirements.

\noindent\resizebox{\columnwidth}{!}{\fbox{%
	\parbox{\columnwidth}{%
		\textbf{Key Findings of RQ3}
		\\
\textbf{Finding 5}:
Three types of methods are used by the participants when seeking architectural information: (i) automatic searching provided in existing tools (e.g., Google, GitHub), (ii) semi-automatic (e.g., Web crawlers), and (iii) manual (e.g., reading offline books).

\textbf{Finding 6}: None of the survey participants used dedicated methods or tools proposed in literature (as discussed in Section \ref{MiningArchInfo}) when searching for architectural information.
	}
}}

\subsection{RQ4: Assessment of the usefulness of architectural information} \label{RQ4_Results}
To answer RQ4, we asked the participants to indicate how they assess the usefulness of retrieved architectural information (SQ14). Table \ref{UsefulnessAssesment} lists the nine measures or factors (ranked according to their counts) reported by the participants. \textit{Well-articulated and relevant architectural information} and \textit{clear description together with architectural diagrams} as the top two major factors they rely on when assessing the usefulness of retrieved architectural information.

\begin{table} [h]
\small
\caption{Measures or factors for assessing the usefulness of architectural information} \label{UsefulnessAssesment}
	\begin{tabular}{p{7.3cm}p{0.6cm}}
		\toprule
	    Measure or factor & Count\\
		\midrule
    Well-articulated and relevant architectural information	& 69\\
    Clear description together with architectural diagrams (e.g., component diagrams) & 62\\
    Complete and comprehensive architectural information  &	59\\
    Architectural information is well-organized and easy to read and understand  &	59\\
    Summarized architectural information	& 53\\
    Categorized architectural information  &	44\\
    Presence of URLs to external sources that support the credibility of architectural information & 42\\
    Documented architectural information   &	38\\
    Validating architectural concepts on more than one source 	&  1\\
	\bottomrule 
	\end{tabular}
\end{table}

\noindent\resizebox{\columnwidth}{!}{\fbox{%
	\parbox{\columnwidth}{%
		\textbf{Key Findings of RQ4}
		\\
\textbf{Finding 7}:
\textit{Well-articulated and relevant architectural information} and \textit{clear description and architectural diagrams} are the top two measures or factors adopted by most of the participants when assessing the usefulness of retrieved architectural information.
	}
}}

\subsection{RQ5: Issues and challenges in searching architectural information} \label{RQ5_Results}
To examine RQ5, we asked our participants to write about their challenges when seeking architectural information (SQ15). The main challenges faced by participants when searching for architectural information are listed and ranked according to their counts in Table \ref{ChallengeWhenSearchingArchInfo}. \textit{Taking too much time to go through architectural information retrieved from various sources} and \textit{feeling overwhelmed due to the dispersion and abundance of architectural information in various sources} were cited by the participants as the top two major issues they face when seeking architectural information in development.

\begin{table} [ht]
\small
\caption{Issues and challenges faced when seeking architectural information} \label{ChallengeWhenSearchingArchInfo}
	\begin{tabular}{p{7.25cm}p{0.6cm}}
		\toprule
	    Issues and challenges & Count\\
		\midrule
    It takes me too much time to go through architectural information retrieved from various sources &	74\\
    I feel overwhelmed due to the dispersion and abundance of architectural information in various sources & 47\\
    Contradictions in architectural information from different sources that make me confused &	44\\
    I sometimes cannot meet the development time due to the time spent on searching for architectural information &	42\\
    Hard to figure out the up-to-dateness in architectural information 	&3\\
    Few sources of architectural information 	&2\\
    Hard to figure out the credibility of architectural information 	&2\\
    Fast pace in the release of architectural information 	&1\\
	\bottomrule 
	\end{tabular}
\end{table}

\noindent\resizebox{\columnwidth}{!}{\fbox{%
	\parbox{\columnwidth}{%
		\textbf{Key Findings of RQ5}
		\\
\textbf{Finding 8}:
\textit{Taking too much time to go through architectural information retrieved from various sources} and \textit{feeling overwhelmed due to the dispersion and abundance of architectural information in various sources} were acknowledged by the participants as the top two major challenges when seeking architectural information in software development.
	}
}}
\section{Discussion}\label{Discussion}
For each RQ, we first describe the representative key findings and then discuss their implications for researchers (indicated with the \faLeanpub \hspace{0.5mm} icon) and/or practitioners (the \faMale \hspace{0.5mm} icon). 

\textbf{RQ1: Sources leveraged to seek architectural information}.
The results of RQ1 (see Figure \ref{SourcesAndParticipants}) show that participants leverage a variety of sources (15 sources) to seek architectural information. Most of the participants prefer to use “\textit{searching engines (e.g., Google)}" to seek architectural information, followed by “\textit{online Q\&A sites (e.g., Stack Overflow)}", “\textit{online repositories (e.g., GitHub)}", and other less frequently used sources. A similar observation has been reported in several studies (e.g., \cite{rahman2018evaluating}, \cite{xia2017developers}) that developers often use search engines to seek information during development. Moreover, we observed that \textit{online Q\&A sites (e.g., Stack Overflow)} and \textit{online repositories (e.g., GitHub)} are present in both top three sources that are frequently visited (see Figure \ref{SourcesAndParticipants}) and top three sources that are considered the most useful (see Figure \ref{MostUsefulSources}) when seeking architectural information. This is in line with the observation in the existing studies (e.g., \cite{bi2021mining} \cite{soliman2016architectural}) that developer forums, like Stack overflow, have shown to contain helpful architectural knowledge. While there are, but a few, studies on the investigation of architectural information in those two sources (Stack Overflow and GitHub), such as architecture smell \cite{tian2020automatic}, architectural patterns and quality attributes \cite{bi2018apqa}, design topics \cite{viviani2018design}, \faLeanpub \hspace{0.5mm} further research is needed, especially in automatic extraction of useful architectural information from those sources to assist development activities.

\faMale \hspace{0.5mm} Practitioners could rely on the list of eleven sources (see Figure \ref{MostUsefulSources}) reported to be the most useful to determine the scope of searching (i.e., to search in specific sites or search in the whole web) when seeking architectural information. For example, practitioners could give a high priority to online repositories if they want to get detailed architectural information in both source code files \cite{gopalakrishnan2017can} and texts (e.g., through architecture design discussions in issues and commits \cite{brunet2014developers} or in pull requests \cite{viviani2018structure}).

\textbf{RQ2: Frequently searched tasks or purposes and tasks or purposes that are difficult to get relevant architectural information}. 
With the results of RQ2, we observed that seeking architectural information \textit{to make architecture design decision among multiple choices} came as the second task that is mostly searched (see Table \ref{PurposesOfsearchingArchInfo}) and also the second task that is difficult to get relevant architectural information (see Table \ref{TasksDifficTofind}). \faLeanpub \hspace{0.5mm} Researchers should pay more attention to this task which is both frequently searched and difficult to get relevant architectural information. For example, researchers might conduct industrial studies (e.g., controlled experiments) to further investigate the causes (e.g., query formulation, query revision, and judging relevance) that limit developers from getting relevant architectural information for that task and develop new approaches that could improve the findability of the relevant architectural information for that task (e.g., either integrated in existing tools or develop new ones).

\textbf{RQ3: Methods and tools used to seek architectural information}.
The survey results indicate that most of the participants, i.e., 79.6\% (82 out of 103), use automatic searching approaches provided in the existing tools (e.g., Q\&A sites, search engines) to seek architectural information (see Section \ref{RQ3_Results}). In addition, while a few dedicated approaches and tools have been reported in the literature (e.g., \cite{soliman2018improving} \cite{bhat2017automatic} \cite{bhat2019adex}), the survey results indicate that none of the respondents had used any dedicated tool or approach proposed in the literature. Hence, \faLeanpub \hspace{0.5mm} more empirical studies are needed to quantify the tangible benefits of those approaches and tools and \faMale \hspace{0.5mm} we encourage practitioners to adopt dedicated approaches and tools for seeking architectural information. Specifically, \faLeanpub \hspace{0.5mm} researchers can answer several questions, such as are these methods and tools easy to use compared with search engines (e.g., Google) and GitHub? What are the factors that prevent developers from using these methods and tools? Are the 17 features (categories of tool requirements proposed in Section \ref{RQ3_Results}) covered by the tools discussed in Section \ref{MiningArchInfo}? In addition, \faLeanpub \hspace{0.5mm} a systematic classification of the mined architectural information with the corresponding methods and tools is needed through conducting a literature review, \faMale \hspace{0.5mm} which can help practitioners to be aware of what architectural information (e.g., tactics) can be mined with what methods or tools to support various development activities.

\textbf{RQ4: Assessing the usefulness of retrieved architectural information}.
Architectural information is useful if it is of informative or practical use to its readers or searchers, i.e., readers or searchers can successfully achieve their goals with the help of retrieved architectural information. Depending on the goals of readers or searchers, usefulness can be characterized by several factors. As shown in Table \ref{UsefulnessAssesment}, the participants have different views when assessing the usefulness of retrieved architectural information. Thus, all the nine factors reported in Table \ref{UsefulnessAssesment}, such as \textit{well described architectural information with architectural diagrams (e.g., component diagrams)}, \textit{presence of URLs to external sources that support the credibility of architectural information}, are of importance in the assessment of the usefulness of architectural information. \faMale \hspace{0.5mm} Practitioners should rely on those factors (see Table \ref{UsefulnessAssesment}) when providing or describing architectural information in their tools (e.g., personal technical blogs and tutorials) or other public sources (e.g., Q\&A sites).

\textbf{RQ5: Issues and challenges faced when searching for architectural information}:
The results in Table \ref{ChallengeWhenSearchingArchInfo} provide us with potential challenges/issues developers face when seeking architectural information. Such issues can stem from many factors, such as the lack of dedicated tool(s) that can aggregate architectural information from diverse sources, which can help developers check the architectural information they need from one tool. Thus, there is a need for \faLeanpub \hspace{0.5mm} researchers and \faMale \hspace{0.5mm} practitioners to address the reported challenges: (i) \faMale \hspace{0.5mm} Practitioners/tool developers can refer to the top two major challenges developers (practitioners) face when searching for architectural information (i.e., taking too much time to go through architectural information retrieved from various sources, and feeling overwhelmed due to the dispersion and abundance of architectural information in various sources (see Table \ref{ChallengeWhenSearchingArchInfo})), develop new tools to handle the performance issues in the exiting tools (e.g., online Q\&A sites, technical blogs, and tutorials), and summarize architectural information retrieved from various sources; (ii) In addition, the participants reported that they got challenges in figuring out the up-to-dateness and credibility of architectural information (see Table \ref{ChallengeWhenSearchingArchInfo}). So \faLeanpub \hspace{0.5mm} researchers could address these issues by developing a framework to assess the up-to-dateness and credibility of the retrieved architectural information from online sources.

\section{Threats to Validity}\label{Threats to Validity}
We discuss the threats to validity of our study by following the guidelines for empirical studies \cite{wohlin2012experimentation}.

\textbf{Internal validity} refers to variables that impact data analysis and results \cite{wohlin2012experimentation}. One typical factor in survey studies is the respondent fatigue bias. We mitigated this threat by limiting the time of completing the survey questionnaire to 15 minutes, which was also verified by running a pilot study with five professional developers to make sure that the survey questions could be answered within 15 minutes. Another possible threat is that 103 participants decided to participate in the survey because they had a greater interest in searching for architectural information than others, hence, that could have introduced a “biased view” of the investigated phenomena due to, for example, the participants did not have suitable experience and expertise to respond to our survey questions. However, our population consists of participants with various roles, years of experience (more than two years, see Table \ref{InclusionAndExcuOfvalidResponses}), and views, for example, views on choosing the sources (see Figure \ref{SourcesAndParticipants}) to seek architectural information and assessing the usefulness of retrieved architectural information (see Table \ref{UsefulnessAssesment}). The last threat to the internal validity in our study is related to the personal bias which could have been introduced by the manual coding process. We alleviated this threat by performing a pilot data extraction and analysis before the formal coding and analysis. Any confusions and disagreements were discussed between the authors of this study till a consensus is reached on the extraction and analysis.

\textbf{Construct validity} concerns the correctness and comprehensibility of the survey questionnaire \cite{wohlin2012experimentation}. In our study, construct threats to validity refer to the phrasing adopted to structure our survey's preliminary definitions, survey questions, and responses for closed-ended questions \cite{ghazi2018survey}. We mitigated these threats by piloting the questionnaire internally several times and thoroughly refining each survey part. Lastly, since the survey comprises several closed-ended questions, participants may not have found any suitable responses in the provided answer options in some questions. We alleviated this risk by (i) defining the set of responses of each survey question through systematically analyzing the evidence reported in the literature (i.e., related work to our study), and (ii) always including a free field “Other” in the set of answers for each survey question.

\textbf{External validity} refers to the generalization of the study results and findings \cite{wohlin2012experimentation}. A relevant threat of this study is related to whether the 103 participants are a good and strong representative of the software developers who seek architectural information in development. As described in Section \ref{SurveyDesign}, we tried to mitigate this threat by sending invitations to software engineers from various sources, such as GitHub and groups of developers on social media sites, to get a representative sample size. Moreover, we applied the snowball sampling approach \cite{kitchenham2002principles} to increase the response rate. However, we are still aware that our used approaches for gathering the participants may have had a negative impact on the size of the set of subjects of our study. We further alleviated this threat by ensuring that the participants are a heterogeneous sample in terms of physical locations (e.g., countries), roles, professional experience, project size, and organization domains (see Section \ref{Demographics}). This indicates a good spread of participants’ profiles and increases the confidence that the study results and findings do not suffer from severe biases with respect to the approaches used to invite the participants.

\textbf{Reliability} concerns the replicability of a study for generating the same or similar results \cite{wohlin2012experimentation}. As elaborated in Section \ref{StudyDesign}, we mitigated this threat by providing a thorough and comprehensive explanation of the process of survey design (e.g., structuring the survey questionnaire, filtering and analyzing the valid data) that can be followed to replicate this survey. Moreover, to guarantee the reliability of our results and findings, a replication package containing the dataset (e.g., valid survey responses) and our encoded data for answering the RQs has been made available \cite{dataset}, allowing other researchers to evaluate the rigor of the design and replicate the study. With these measures, this threat has been partially reduced.

\section{Conclusions}\label{Conclusions and Future Work}
 
In this study, we surveyed 103 participants across 6 continents (i.e., 40 countries) to understand architectural information search activities. The findings of this study can be used to provide concrete guidelines for practitioners to refer to when seeking architectural information and providing architectural information that could be considered useful, and highlight the difficulties and challenges in searching for architectural information that researchers can investigate. 

We found that participants leverage various sources, such as technical websites (e.g., InfoQ), technical blogs and tutorials, specific technology forums (e.g., Apple developer forums) to seek architectural information. The survey results indicate that dedicated approaches and tools (proposed by researchers in the literature) are not adopted in practices. When seeking architectural information, participants find difficulties mostly in getting relevant architectural information for addressing quality attribute concerns, followed by making design decisions among multiple choices. Taking too much time to go through architectural information retrieved from various sources and feeling overwhelmed due to the dispersion and abundance of architectural information in various sources are the top two major challenges the participants face when seeking architectural information. 

In the next step, we plan to design and employ (semi-)automatic approaches to extract and summarize architectural information and establish the issue-solution pairs from the retrieved architectural information, for example, benefits and drawbacks of certain architectural solutions (e.g., patterns and tactics) for task-specific architecture design problems from multiple sources of architectural information (e.g., Q\&A sites, online repositories, technical blogs), which can facilitate the decision-making of architects by utilizing architectural knowledge from peers.



\balance
\bibliographystyle{IEEEtran}
\bibliography{ref}

\end{document}